\documentclass[aps,nofootinbib,showpacs,twocolumn,
floatfix,superscriptaddress]{revtex4}
\usepackage{graphicx}
\usepackage{epsf,amsmath,amsfonts,amssymb,amsbsy}
\usepackage[mathscr]{eucal}
\usepackage{colordvi}
\begin{document}

\title{Constructing the ultimate theory of grand unification}
\author{V.V.Kiselev}
\email{Valery.Kiselev@ihep.ru}
 \affiliation{Russian State Research
Center ``Institute for High
Energy Physics'', %\\
Pobeda 1, Protvino, Moscow Region, 142281, Russia\\ Fax: +7-4967-744937}
 \affiliation{Moscow Institute of Physics and
Technology, Institutskii per. 9, Dolgoprudnyi, Moscow Region, 141701, Russia}
\author{V.V.Shakirov}
\affiliation{Moscow Institute of Physics and Technology, Institutskii per. 9,
Dolgoprudnyi, Moscow Region, 141701, Russia}
\begin{abstract}
In accordance with known phenomenological facts on leptons and quarks in the
Standard Model as well as on the scale of neutrino masses and introducing the
supersymmetry, we logically substantiate the unique composition of
fundamental representation for the fermionic multiplet of gauge group $E_8$.
\end{abstract}
\maketitle

\section{Introduction}
In the framework of local quantum field theory the idea of grand unification
of gauge interactions (GUT) is logically based on three facts
\cite{Georgi:1974sy,Nakamura:2010zzi}:

\begin{enumerate}
  \item the known fermions are chiral fields,
  \item the sum of electric charges of those fermions (quarks and
      leptons) in each generation is equal to zero,
  \item the renormalization group evolution of coupling constants for the
      electroweak and strong interactions (the gauge group of Standard
      Model (SM) is $G_\mathrm{SM}=U_Y(1)\times SU(2)_L\times SU_c(3)$)
      leads to the almost precise equality of those three ``running''
      charges to each other at the scale of ``unification''
      $\Lambda_\mathrm{GUT}$ about $10^{16}$ GeV, above which the
      propagation of heavy fields becomes essential and, hence, the
      underlying gauge symmetry is restored with the single coupling
      constant.
\end{enumerate}

According to the structure of gauge group $G_\mathrm{SM}$ the actual fields
have got 4 independent quantum numbers: the hypercharge $Y$, the projection
of electroweak isospin for the left doublets and right singlets $T_3$, and
the colored analogues of ``isospin'' $\lambda_3^c$ and ``strangeness''
$\lambda_8^c$, that sets the rang of group equal to 4. Under the grand
unification the amount of quantum numbers cannot decrease, so that the rang
of GUT group cannot be less than 4. However, the introduction of common
coupling constant of interactions suggests that the GUT group is simple,
therefore, the sum of quantum numbers in the multiplet has to be equal to
zero (the trace of generator in the simple nonabelian group equals zero),
that exactly makes the property 2 pointed out above, while in accordance with
item 1 in this scheme, the group itself should allow for non-real
(non-selfconjugative) representations corresponding to chiral fields.

In this way the ratios of charges in the multiplet are completely determined
by the structure of group, hence, there is the charge quantization, for
instance, the quantization of electric charge of fermions. Emphasize that
there should exist an irreducible representation, which includes the known
charged fermions only, because the addition of new charged fermions to such
the representation would make the arbitrariness in the quantization of
electric charges, if only the sum of charges of additional fermions would not
be equal to zero. In the later case, we could conclude that the additional
fermions compose an irreducible representation of the same group, which
produces the quantization of charges for the known fermions. In this respect,
a simple group of minimal admissible rang should regard to the quantization
of charges for the chiral fields of fermions in the SM, and the problem to
find such the group was actual. This problem has been successfully solved in
\cite{Georgi:1974sy}: the group of charge quantization is $SU(5)$, and it has
got the minimally admissible rang equal to 4, while the known fermions of SM
compose two fundamental representations, the antiquintet and decuplet of left
chiral fields\footnote{In this case the hypercharge $Y$ is proportional to
one of Cartan generator in the group $SU(5)$, while a coefficient can be
easily calculated from the common normalization of group generators, i.e. by
the same sum of charge squares in the multiplet, that is essential in the
derivation of statement on the equality of ``running'' coupling constants at
the scale of unification (the main content of item 3 in the list above).}
(the formulation with two conjugated fundamental representations is
equivalent to the given one).

In Section \ref{sec-1} we recall the properties of primary set for the chiral
fields of SM and expand the action of charge quantization-group to the Higgs
sector of SM, which gives the natural introduction of Dirac masses for quarks
and charged leptons under the spontaneous breaking  the symmetry by the
vacuum state of scalar field. Further, we involve the supersymmetry, that
makes the addition of chiral superpartners of Higgs fields into the multiplet
of representation for the GUT group to be necessary. The charges of these
superpartners are uniquely quantized. Next, we argue for the necessity of
introducing a special mechanism, which should generate the required scale for
the neutrino masses as follows from the phenomenology: the only way to relate
the scale of electroweak symmetry breaking to the neutrino masses is the
mixing of electroweak singlet neutrino with a neutral heavy Majorana
particle, that completes the composition of multiplet of gauge group for the
single generation, i.e. the fundamental representation of group $E_6$ with
the unique determination of quantum numbers for all of components.

In Section \ref{sec-2} we consider $n_g$ generations for the chiral fermions
of group $E_6$ as the components of multiplet unified with the Majorana
superpartners of gauge vector fields in $E_6$, i.e. with the gaugino. In this
way, the supersymmetry requires to include $n_g$ conjugated fundamental
respresentations of $E_6$ for the chiral fields into such the common
multiplet in order to make the unification with the Majorana gaugino from the
adjoint representation of $E_6$. Then, those fields can be unified in the
same extended selfconjugated representation. Moreover, the equivalence of
generations for the chiral fields means, in fact, the introduction of unitary
transformations of generations themselves. Therefore, we have to include the
``horizontal'' symmetry of generations into the ultimate group of GUT. The
Majorana superpartners of gauge vector bosons in $SU_H(n_g)$ as well as the
gauge bosons of horizontal symmetry are singlets with respect to the group
$E_6$, and they should be added into the complete multiplet of fermions, too.
In this step, we have to take into account for that the gauginos of $E_6$ are
singlets with respect to the group of horizontal transformations for the
fermion generations. Fortunately, the fermionic multiplet with the required
properties does exist, and it is unique, whereas $n_g=3$, while the group of
symmetry is the exceptional group $E_8$. Finally, we discuss some
consequences following from the formulation of ultimate theory of grand
unification as well as its properties.

On the other hand, in the framework of superstring theory the phenomenology
of Yukawa sector in the SM points to the emphasis of specific way for the
compactification of extra dimensions, namely, the F-theory
\cite{Beasley:2008dc,Beasley:2008kw,Heckman:2010bq} with two hierarchical
scales of energy at $\Lambda\ll M$, where $M$ is the stringent Planckian
scale, while $\Lambda$ is the GUT scale. In this scenario, the unique role of
group $E_8$ is also noticed \cite{Heckman:2009mn}.

Thus, to our point of view, we successfully construct the only true solution
for the problem of finding the group of grand unification as well as its
fermionic multiplet by starting from the given data of phenomenology and by
involving the supersymmetry. This solution gets the common with the mentioned
scheme of superstring model, but, as we will see, the proposed way
essentially differs from the superstring models, since the states with
Planckian scale of masses are included in the fundamental multiplet of gauge
group in contrast to multiplets inspired by the superstring theory, when the
multiplet is composed of light sub-Planckian states. In this study we put the
mechanism of breaking both the gauge symmetry and the supersymmetry beyond
the current consideration.

\section{The single generation\label{sec-1}}
Following the principles listen in Introduction, we are considering the
logical construction of multiplet for the single generation of chiral
fermionic fields in the SM.

\subsection{The Standard Model: chiral fermions and Higgs scalar}

The observed leptons and quarks of SM compose the set of left-handed chiral
spinors of weak doublets and singlets over the group $SU(2)_L$:
\begin{equation}\label{SM-chiral}
    \begin{array}{ccc}
       \left(\hskip-1.mm
          \begin{array}{c}
            \nu \\
            e \\
          \end{array}\hskip-0.8mm
        \right)_{\hskip-1.mm L},
        & e^c_L, &  \\[4mm]
       \left(\hskip-1.2mm
         \begin{array}{c}
           u \\
           d \\
         \end{array}\hskip-0.8mm
       \right)_{\hskip-1.2mm L},
        & u^c_L, & d^c_L,
     \end{array}
\end{equation}
where the superscript ``c'' refers to the charged conjugated particles, i.e.
antiparticles. The scalar Higgs field is the weak doublet:
\begin{equation}\label{SM-higgs}
     \left(\hskip-1.2mm
          \begin{array}{c}
            H^0 \\
            H^- \\
          \end{array}\hskip-1.2mm
        \right),
\end{equation}
which has got zero leptonic and baryonic numbers
$$B[H]=L[H]=0.$$
The masses of Dirac kind for the charged leptons and quarks are given by
matrices for the chiral spinors, so that, for instance, in the case of
electrons
\begin{equation}\label{el-mass}
    (e, e^c)_L \left(
                   \begin{array}{cc}
                     0 & m_e \\
                     m_e & 0 \\
                   \end{array}
                 \right)
    \left(\hskip-1mm
      \begin{array}{c}
        e \\
        e^c \\
      \end{array}\hskip-1mm
    \right)_{\hskip-1mm L},
\end{equation}
where the two-component indices of chiral spinors are convoluted by means of
spinor metric being the completely antisymmetric tensor of second rang, i.e.
the Levi-Civita tensor. In this way, the masses themselves are generated by
the spontaneous breaking of initial gauge group $G_\mathrm{SM}$ due to both
the vacuum expectation value of neutral component of Higgs field $\langle
H^0\rangle=v/\sqrt{2}$ and the Yukawa-contact, gauge-invariant interaction of
Higgs scalar with the fermions, so that $m_e=\lambda_e \langle H^0\rangle$.

It is important to note that the observed hierarchy of masses for three known
generations of fermions (two generations are much lighter than the third
generation) can be naturally explained by the \textit{single} scale of
energy, i.e by the vacuum expectation value of Higgs scalar. Indeed, the
introduction of symmetric, so called ``democratic'' matrix for the masses of
Dirac spinors $\psi_\alpha$ distinguished only by the marker of generation
$\alpha=\{1,2,3\}$ into the corresponding term of lagrangian $\mathscr{L}_M$
with the unified coupling constant $\lambda$ in the form of
\begin{equation}\label{gen-mass}
    \mathscr{M}_D=\lambda\, \langle H^0\rangle
    \left(
                  \begin{array}{ccc}
                    1 & 1 & 1 \\
                    1 & 1 & 1 \\
                    1 & 1 & 1 \\
                  \end{array}
                \right)\;\mapsto\;
                \mathscr{L}_M=\bar\psi\, \mathscr{M}_D\psi,
\end{equation}
leads to two massless generations and the single heavy generation, while
small corrections to the democracy of generations cause the opportunity to
describe the mass spectrum of charged leptons and quarks by making use of
``seesaw'' mechanism \cite{Fritzsch}, i.e. by a small mixing of light states
with the heavy state, but, importantly, without any introduction of new
physical scales relevant to the differences of masses between the electron,
muon and $\tau$-lepton, for instance.

In the case of quarks, those corrections to the mass matrix of ``democratic''
kind cause the formation of matrix for the mixing of charged weak currents.
i.e. the Cabibbo--Kobayashi--Maskawa matrix \cite{KM}, too. Hence, this
matrix of current mixing has also got the hierarchy, since it is formed by
matrices of transitions from the initial basis of quark fields to the basis
of physical massive states for the generations with different values of weak
isospin. The transition matrices carry all of mentioned properties of
symmetry for the generation in the Yukawa sector of theory. Therefore, we do
not require the introduction of additional scales except the vacuum
expectation value of Higgs field in the case of masses for charged fermions.

In respect of masses, the situation is crucially changed if we consider the
phenomenology of neutrino, since the maximal mass of neutrinos is restricted
by the constraint $m_\nu< 0.2$ eV \cite{Nakamura:2010zzi}. Therefore, the
analogous mechanism for the generation of neutrino masses suggests that the
initial Yukawa coupling constant in the matrix of Dirac masses of democratic
kind should be twelve orders of magnitude less than the constant for the
charged leptons, that evidently points to the essential difference in the
mechanisms of mass generation for the charged and neutral
particles\footnote{The difference of Yukawa coupling constants in the
democratic symmetry for generations of quarks and charged leptons is
restricted by two orders of magnitude, that can be certainly explained by the
renormalization group evolution of those constants from the scale of true
``democracy'', when the constants have got the same order of magnitude, to
the scale of breaking down the electroweak symmetry, since the
renormalization group equations depend on the quantum numbers, which are
different for the quark and charged leptons because of different values of
weak isospin projection.}. Thus, the neutrinos involve the necessity of
modifying the theory of mass generation by introducing an additional scale of
energy\footnote{The single or a few.}, whereas the naive introduction of low
energy scale is in contradiction with the phenomenology, because, except the
neutrino masses, we do not observe any unusual phenomena beyond the SM at
energies much less than the scale of electroweak symmetry breaking, that
witnesses on the fact that the low scale is not related to the dynamics at
low energies, but it is reducible from the dynamics at high energy scales,
which is not yet observed at the energies about the vacuum expectation value
of Higgs field. Nevertheless, the ``seesaw'' mechanism allows us to make the
step to the derivation of GUT structure in this sector, too (see Section
\ref{subsec-nu}). However, in this way, the relation of masses with the
vacuum expectation values suggests the introduction of additional neutral
field with a new vacuum expectation value, that will be explicitly done
below.

\subsection{The charge quantization}
As was mentioned in Introduction, the solution of electric charge
quantization for the fermionic fields of SM is the unification group $SU(5)$,
whereas the fields are decomposed into fundamental representations, so that
the antiquintet $\boldsymbol {\bar 5}$ is given by
\begin{equation}\label{su5-anti-quint}
    \left(\hskip-1.2mm
      \begin{array}{c}
        \,
      \;d^c \\
        \nu \\
        e \\
      \end{array}
    \hskip-1.2mm\right)_{\hskip-1.2mm L},
\end{equation}
while the decuplet $\boldsymbol {10}$ is composed of $u, d, u^c, e^c$ in the
form of antisymmetric matrix $5\times 5$ (see the review on the group theory
in \cite{Slansky:1981yr}).

From the form of antiquintet we would see that the vector leptoquarks
transforming the leptons into the $d^c$-antiquarks, are the part of gauge
bosons of $SU(5)$. If we assume that the same group is responsible for the
charge quantization of scalar Higgs bosons, then we arrive to the conclusion
that the Higgs scalars necessary compose the following antiquintet
$\boldsymbol {\bar 5}$:
\begin{equation}\label{su5-higgs}
    \left(\hskip-1.2mm
      \begin{array}{c}
        \mathcal{K}^c \\
        H^0 \\
        H^- \\
      \end{array}
    \hskip-1.2mm\right),
\end{equation}
with the color antitriplet-scalar leptoquark $\mathcal{K}^c$, possessing the
charge $Q=\frac13$, the lepton number $L=-1$ and the baryon number
$B=-\frac13$, because this scalar antiquintet interacts with the same gauge
vector bosons as the chiral antiquintet does\footnote{If one does not use the
argumentation on the transitions between the components of fermionic and
scalar antiquintet due to the emission of gauge bosons with identical quantum
numbers, then the triplet field $\mathcal{K}$ can get other values of lepton
and baryon numbers (see review \cite{E6}).}.

\subsection{The supersymmetry}
We can quite naturally suggest that the ultimate theory of grand unification
is supersymmetric. Then the fermions and Higgs bosons are components of
chiral superfields, including corresponding superpartners for the particles
of SM (see the textbook \cite{Weinberg-VIII}). A new quantum number is the
$R$-parity, it marks the particles and their superpartners: the particles are
assigned to positive parity $R=+1$, while the superpartners have got the
negative $R$-parity.

In terms of chiral superfields the supersymmetric lagrangian of interactions
is constructed as the polynomials of third power, whereas the mass term for
the fields with the positive and negative projections of weak isospin appears
only after the introduction of two Higgs scalars with conjugated quantum
numbers,
\begin{equation}\label{su5-higgs-susy}
    \left(\hskip-1.2mm
      \begin{array}{c}
        \mathcal{K}^c \\
        H^0_u \\
       ~H^- \\
      \end{array}
    \hskip-1.2mm\right),\quad
    \left(\hskip-1.2mm
      \begin{array}{c}
        \mathcal{K} \\
        ~H^+ \\
        H^0_d \\
      \end{array}
    \hskip-1.2mm\right).
\end{equation}
This fact is the consequence of supersymmetry of lagrangian, that leads to
doubling the number of Higgs scalars. The doubling is also necessary for the
cancelation of triangle quantum anomaly for the chiral currents: the
introduction of fermionic superpartners for the single Higgs boson of SM
would produce the additional contribution of this superpartner into the
triangle diagram, and, hence, to the quantum anomaly, while the conjugated
superpartner cancels the anomaly.

Thus, the chiral sector should be extended by addition of superpartners for
the Higgs particles, i.e. due to the antiquintet  $\boldsymbol{\bar 5}$ and
quintet $\boldsymbol 5$ with respect to $SU(5)$:
\begin{equation}\label{susy-higgs}
    \left(\hskip-1.2mm
      \begin{array}{l}
        \varkappa^c \\
        \chi^0_u \\
        \chi^- \\
      \end{array}
    \hskip-1.2mm\right)_{\hskip-1mm L}\quad\mbox{ и }\quad
    \left(\hskip-1.2mm
      \begin{array}{l}
        \varkappa \\
        \chi^+ \\
        \chi^0_d \\
      \end{array}
    \hskip-1.2mm\right)_{\hskip-1mm L},
\end{equation}
with the negative $R$-parity.

\subsection{Neutrino masses \label{subsec-nu}}
As was mentioned in Introduction, the observed scale of neutrino masses is
not truly fundamental, this scale is reducible from factors given
dynamically, hence, the neutrino masses can be calculated in terms of high
energy-scales and their ratios. In this way, the primary scales of energy
originate from the dynamics of gauge theory.

So, the mass of Dirac kind $m_D$ is related to the breaking of electroweak
symmetry, and $m_D\sim v\sim 10^2$ GeV, so that this kind of mass breaks the
weak isospin, but it conserves both the lepton and baryon numbers, as well as
the $R$-parity. Therefore, neutrinos could get the masses of Dirac kind under
the introduction of electroweak left-handed singlet antineutrino $\nu^c_L$
(the leptonic number is $L=-1$, the $R$-parity is equal to $R=+1$ as for the
ordinary matter, and the baryon number is equal to zero $B=0$, of course).
However, the single Dirac contribution to the neutrino mass is not enough in
order to get the necessary scale of neutrino masses.

The mass term of Majorana kind suggests the conservation of electroweak
symmetry due to the zero values of electroweak charges of Majorana particle,
i.e. the Majorana component of $\nu_S$ should be the singlet with respect to
the group of electroweak symmetry. Moreover, the term quadratic in $\nu_S$
has to conserve both the lepton and baryon numbers. Therefore, the Majorana
component is the sterile particle with zero values of lepton and baryon
numbers (sterino), while, under the condition of $R$-parity conservation, its
coupling to the neutrinos $\{\nu_L,\nu^c_L\}$ leads to $R[\nu_S]=+1$ and
gives the zero constants of mixing with the neutral components of Higgs
superpartners $\chi^0_u$ and $\chi^0_d$, because their $R$-parity is
negative.

Therefore, the mass matrix of neutrinos is the matrix of $3\times 3$ for the
components $\{\nu_L,\nu^c_L, \nu_S\}$, and it take the form of
\begin{equation}\label{M-nu}
    \mathscr{M}=\left(
                  \begin{array}{ccc}
                    0 & m_D & 0 \\
                    m_D & 0 & \Lambda \\
                    0 & \Lambda & M \\
                  \end{array}
                \right),
\end{equation}
where $m_D$ is the mass of Dirac kind, the Majorana mass is denoted by symbol
$M$, while the scale $\Lambda$ determines the mixing between the electroweak
singlet antineutrino $\nu^c_L$ and sterino $\nu_S$.

This mechanism suggests the explicit breaking of lepton number\footnote{If we
set the nonzero value for the lepton number of sterino as for the left-handed
electroweak doublet with neutrino, then the mixing of sterino with the
singlet antineutrino would conserve the lepton number, but the lepton number
would be broken by the introduction of Majorana mass term of sterino, which
would be not the true Majorana particle, in this case.} at the scale of
$\Lambda$, therefore, we could naturally put
$\Lambda\sim\Lambda_\mathrm{GUT}$, because the grand unification of gauge
interactions puts leptons and quarks into a common multiplet and makes them
indistinguishable under the gauge transformations, hence, it is the most
probable that breaking the lepton number is related to the dynamical scale of
breaking the symmetry of grand unification.

As for the term of Majorana mass with the scale of $M$, it does not generate
any breaking of quantum numbers entering the charges of SM and marking the
leptons and baryons. This fact can point to the relation with more universal
dynamical characteristics, such as the breaking the local supersymmetry, i.e.
the supergravity. Therefore, in the way of working hypothesis we could set
that $M$ gets the scale characteristic for the gravity, i.e. it is of the
order of Planckian  mass $m_\mathrm{Pl}$. This suggestion means that the
mechanism generating the low scale of neutrino masses leads to the theory of
grand unification, which essentially differs from models inspired by the
superstring theory, when the fermionic multiplet of GUT is composed of
particles with masses of the GUT scale or scales mush less than it, because
one considers the particles in the low energy spectrum, while the fields of
Planckian scale are referred to the spectrum of excitations of superstring.
In our construction the sterino as well as some other fields, which would be
introduced below, explicitly belong to the sector of particles with the
Planckian scale of mass.

The eigen values of matrix (\ref{M-nu}) as well as the corresponding quantum
states can be easily calculated, if we suggest the hierarchy of scales $m_D$,
$\Lambda$ and $M$. Indeed, setting  the parameters of matrix to be real, we
perform, first, the rotation for two heavy states by the angle $\theta_{23}$,
satisfying the condition of
\begin{equation}\label{t23}
    \tan 2\theta_{23}=\frac{2\Lambda}{M},
\end{equation}
under the action of matrix
\begin{equation}\label{U23}
    U_{23}=\left(
             \begin{array}{ccc}
               1 & 0 & 0 \\
               0 & ~~c_{23} & s_{23} \\
               0 & -s_{23} & c_{23} \\
             \end{array}
           \right),
\end{equation}
where $c_{23}=\cos\theta_{23}$, $s_{23}=\sin\theta_{23}$, so that the mass
matrix gets the form of
\begin{equation}\label{M23}
    \mathscr{M}_{23}=U^\dagger_{23}\cdot\mathscr{M}\cdot U_{23}\approx\hskip-3pt
    \left(\hskip-3pt
      \begin{array}{ccc}
        0 & m_D & m_D\,\frac{\Lambda}{M} \\[2mm]
        m_D & -\Lambda\,\frac{\Lambda}{M} & 0 \\[2mm]
        m_D\,\frac{\Lambda}{M} & 0 & M+\Lambda\,\frac{\Lambda}{M} \\
      \end{array}   \hskip-3pt
    \right)\hskip-2pt,
\end{equation}
where we have used the smallness of ratio $\Lambda/M\ll1$, hence,
$$
    \sin\theta_{23}\approx \frac{\Lambda}{M},\qquad
    \cos\theta_{23}\approx 1-\frac{\Lambda^2}{2M^2}.
$$
Matrix (\ref{M23}) allows for the consideration in the framework of
stationary perturbation theory, so that under taking into account for the
mass hierarchy and after the definition of composite scale
$\Lambda_X=\Lambda^2/M\gg m_D$ we get
\begin{equation}\label{masses}
    m_1\approx \frac{m_D^2}{\Lambda_X},\quad
    m_2\approx \Lambda_X-m_1,\quad
    m_3\approx M-\Lambda_X,
\end{equation}
where we especially show the leading corrections to the masses of heavy
neutrinos in order to demonstrate the valid relation of matrix trace to its
initial value $\mbox{tr}\mathscr{M}=M$. There is the hierarchy of neutrino
masses in the following form:
$$
    m_1\ll m_2\ll m_3.
$$
In the calculation scheme for the eigen values of mass matrix of neutrinos,
we can easily find expressions for the eigen states. So,
\begin{equation}\label{n3}
\begin{array}{l}
\displaystyle
    |\nu_3\rangle\approx s_{23}|\nu_L^c\rangle+c_{23}|\nu_S\rangle+
    \frac{m_D\Lambda}{M^2}\,|\nu_L\rangle,\\[2mm]
\displaystyle
    |\nu_2\rangle\approx c_{23}|\nu_L^c\rangle-s_{23}|\nu_S\rangle-
    \frac{m_D}{\Lambda_X}\,|\nu_L\rangle,\\[2mm]
\displaystyle
    |\nu_1\rangle\approx |\nu_L\rangle+\frac{m_D}{\Lambda_X}\,
    (c_{23}|\nu_L^c\rangle-s_{23}|\nu_S\rangle).
\end{array}
\end{equation}
Therefore, the lightest neutrino is practically left-handed neutrino of SM,
but it gets the admixture of both the left-handed antineutrino and the
sterino.

For the numerical estimates we put $m_D\sim 10^2$ GeV, as expected in
accordance with the known scale of electroweak symmetry breaking, and we
accept $m_1\sim 10^{-2}$ eV for the neutrino mass. Then, we find
$\Lambda_X\sim 10^{15}$ GeV. Therefore, at $M\sim 10^{19}$ GeV we get
$\Lambda\sim 10^{17}$ GeV, that completely agrees with the characteristic
scale of GUT breaking. Thus, the breaking of lepton number in the expression
for the massive neutrino (\ref{n3}) has got the amplitude of the order of
$10^{-13}$.

\subsection{The vacuum}
By the spontaneous breaking of gauge symmetry the masses of chiral fermions
are generated due to the vacuum expectation values of scalar fields, which
appear for the electrically neutral fields, only, in the case of electrical
charge conservation we consider. In this way, the vacuum can carry nonzero
charges, which conservation is broken. If the $R$-parity is conserved, then
\begin{equation}\label{vacuum}
    Q[\mbox{vac}]=L[\mbox{vac}]=B[\mbox{vac}]=0,\quad R[\mbox{vac}]=+1,
\end{equation}
while the weak isospin $T_3[\mbox{vac}]=\pm\frac12$ is broken due to the
condensates of Higgs fields $\langle H^0_u\rangle=v_u$ and $\langle
H^0_d\rangle=v_d$.

Further, the scalar superpartners of doublet neutrino, singlet antineutrino
and sterino, i.e. the particles $\{\tilde \nu_L,\,\tilde
\nu^c_L,\,\tilde\nu_S\}$ are neutral. The sterile scale of $M$ could be
assigned to the vacuum expectation value of terino
$\langle\tilde\nu_S\rangle$. This scale enters into the mass matrix of
neutrinos, but the terino being the superpartner of ordinary particle gets
the negative $R$-parity, hence, within the paradigm of spontaneous symmetry
breaking the mass matrix of neutrinos suggests the introduction of a field
conjugated to the sterino and its superpartner with respect to the
$R$-parity. Therefore, we have to introduce the conjugated multiplet for the
complete set of fields of single generation, if we assume the conservation of
$R$-parity. The conjugated multiplet is called the mirror matter.

The vacuum expectation value of anti-sneutrino $\langle\tilde\nu^c_L\rangle$
breaks down the lepton number, only, and it corresponds to the scale of
$\Lambda$ with the same notes about the $R$-parity as in the previous
paragraph.

The introduction of nonzero vacuum expectation value for the sneutrino
$\langle\tilde\nu_L\rangle$ would break both the weak isospin and the lepton
number, too. It would lead, in general, to the matrix element mixing the
sterino with the neutrino $\mathscr{M}_{13}$ (and $\mathscr{M}_{31}$) in
(\ref{M-nu}). For the sake of simplicity and definiteness we consider the
variant with zero vacuum expectation value of sneutrino, though its
introduction would mean the existence of the additional source for the
breaking of electroweak symmetry, that does not exhibit itself at low
energies.

The listing of admissible vacuum expectation values allows us to make a
question on the masses of Higgs particle superpartners: it is clear that the
triplet higgsinos $\{\varkappa,\varkappa^c\}$ can be supermassive due to the
scale of $M$, only, because their mass term does not break down both the
weakisospin and the lepton number. It seems that the introduction of masses
for the doublet higgsinos would follow the same argumentations, but then it
would be problematic to introduce the scale for the electroweak symmetry
breaking, that is inherently related to the scale of mixing the neutral
fields $\{H^0_u,H^0_d\}$ and their superpartners due to the $\mu$-term in the
lagrangian of superfields in the case of supersymmetry. This is the well
known problem of doublet-triplet splitting for the Higgs fields. The problem
is reduced to the hierarchy of scales, and it is essential for the proton
lifetime, since the light triplet leptoquarks of Higgs sector are excluded by
experimental constraints on the proton decay rate. However, the mechanism of
generating the hierarchy of scales for the breaking of electroweak symmetry
and gauge group of GUT, $v\ll \Lambda$, can be naturally invented in the case
of several generations of fermions as would be discussed in Section
\ref{sec-2}.

\subsection{The multiplet of $E_6$}

Thus, summing up the components necessary for the construction of chiral
multiplet of single generation as described above, we get the
$\boldsymbol{27}$-plet with the quantized charges:
\begin{enumerate}
  \item leptons $e_L$, $e^c_L$, $\nu_L$, $\nu^c_L$ (4 components),
  \item sterino $\nu_S$ (1 component),
  \item color-triplet leptoquark higgsino $\varkappa$, weak
      isospin-doublets $\chi^0_u$, $\chi^-$ and the corresponding
      conjugated higgsinos (10 components),
  \item colored quarks $u_L$, $u^c_L$, $d_L$, $d^c_L$ (12 components).
\end{enumerate}
Such the set exactly and uniquely corresponds to the irreducible fundamental
representation of group $E_6$ \cite{Slansky:1981yr}. Significantly, the group
has no chiral quantum anomalies.

\section{Generations \label{sec-2}}

The further unification of fermions into the multiplet of ultimate theory of
grand unification has to account for the existence of several generations of
fermions in SM, as well as for the superpartners of gauge bosons of group
$E_6$, i.e. gauginos.

\subsection{Gauginos of $E_6$}

In the supersymmetric theory, the gauginos are transformed under the adjoint
representation of symmetry group and they are Majorana particles. Therefore,
for the group of $E_6$ there is the $\boldsymbol{78}$-plet of gauginos
\cite{Slansky:1981yr}. Then, the unification of gauginos with the chiral
fermions of fundamental $E_6$-representation, i.e. with the
$\boldsymbol{27}$-plet, into the common multiplet requires the introduction
of conjugated chiral fundamental representation of $E_6$, i.e.
$\boldsymbol{\overline{27}}$-plet. Certainly, such the extension permits the
self-conjugation of unified multiplet. The conjugated fundamental multiplet
of $E_6$ is called the ``mirror'' matter. Each the generation of SM is
associated with the generation of mirror matter in the ultimate GUT.

\subsection{The horizontal symmetry of generations}

If we temporary miss the Yukawa sector responsible for the introduction of
fermion masses and current mixing, then the observed fermion generations
become equivalent to each other with respect to the transformations of gauge
group. Therefore, we can introduce the gauge symmetry of generations, that
unitary mixes the generations, i.e. the ``horizontal'' group of symmetry, the
group of unitary rotations of generations ${SU}_H(n_g)$, where $n_g$ is the
number of generations. It transforms the fundamental representations of
$E_6$. The actual number of generations is not less than 3.

The gauge bosons of horizontal symmetry are singlets with respect to the
group $E_6$. The Majorana superpartners of those gauge bosons should be
included into the fermionic multiplet of ultimate GUT. The number of gauge
bosons of horizontal symmetry is equal to $n_g^2-1$.

Finally, the gauginos of $E_6$ do not carry any quantum numbers of
generations, i.e. they are singlets with respect to the group of horizontal;
symmetry.

Thus, the self-conjugated multiplet of ultimate GUT includes
\begin{itemize}
\item  $n_g$ fundamental representations of $E_6$ and $n_g$ conjugated
    representations of mirror matter,
\item $n_g^2-1$ self-conjugated gauginos of horizontal symmetry as
    singlets with respect to $E_6$,
\item gauginos of $E_6$ being singlets with respect to the horizontal
    symmetry.
\end{itemize}
Then, the expansion of multiplet under the direct product of groups for the
horizontal symmetry and the symmetry of single generation $SU_H(n_g)\times
E_6$ should get the form
$$
    (\boldsymbol{n}_g,\boldsymbol{27})+(\boldsymbol{\bar n}_g,
    \boldsymbol{\overline{27}})+(\boldsymbol{1},\boldsymbol{78})+
    (\boldsymbol{n_\mathit{g}^2-1},\boldsymbol{1}).
$$
It is interesting that there is the unique simple group with the required
expansion of irreducible multiplet at $n_g>1$. This is the exceptional group
$E_8$ \cite{Slansky:1981yr}, whereas $n_g=3$, so that
$\mathrm{E}_8\supset\mathrm{SU}(3)\times \mathrm{E}_6$, while the fundamental
representation of minimal dimension is the self-conjugated
$\boldsymbol{248}$-plet \cite{Slansky:1981yr}:
\begin{equation}\label{dec-E_8}
    \boldsymbol{248}=(\boldsymbol{3},\boldsymbol{27})+
    (\boldsymbol{\bar 3},\boldsymbol{\overline{27}})+
    (\boldsymbol{1},\boldsymbol{78})+(\boldsymbol{8},\boldsymbol{1}).
\end{equation}

Therefore, the offered logics for the construction of ultimate GUT leads to
the unique result, the group $E_8$. The inherent attribute of such the
construction is the introduction of mirror generations of matter and Higgs
particles. In this way, the condition of $R$-parity conservation can result
in that the mass terms in the sector of matter are generated due to the
vacuum expectation values of scalar particles in the mirror world, that does
not change the justification of our approach to the construction of fermionic
multiplet in GUT. In addition, we get the problem of decoupling for the
mirror matter, probably being superheavy, that is shortly considered in the
end of next subsection.

\subsection{The mechanism of hierarchy}
The existence of several generations for the Higgs superfields allows us
naturally to introduce the hierarchy of scales for the breaking of
electroweak and GUT symmetries in the same manner as was done for the
hierarchy of masses for the fermion generations: the mass matrix of doublet
components of Higgs particles can take the form corresponding to the
``democracy'' of Higgs particle generations. Therefore, the only generation
would be supermassive, i.e. it would get the mass of the order of Planck
scale $M$. For instance, the mixing given by the single scale $M$ for three
generations of neutral higgsinos could be written in the form
\begin{equation}\label{mix-chiggs}
    \mathscr{L}_\mathrm{ud}=\lambda_\mathrm{ud}\hskip-2.pt\cdot\hskip-2.pt
    (\chi^0_{u1},\chi^0_{u2},\chi^0_{u3})\hskip-2.pt
    \left(
                  \begin{array}{ccc}
                    M & M & M \\
                    M & M & M \\
                    M & M & M \\
                  \end{array}
                \right)\hskip-4.pt
    \left(\hskip-2.pt
    \begin{array}{c}
      \chi^0_{d1} \\[1.2mm]
      \chi^0_{d2} \\[1.2mm]
      \chi^0_{d3}
    \end{array} \hskip-2.pt
    \right),
\end{equation}
that is supersymmetrically reflected in the mass matrix for the scalar Higgs
particles.

Two lightest generations of Higgs doublets would get non-zero
$\mu$-terms\footnote{The contribution to the superpotential in the form of
$\mu\cdot\tilde H_u \tilde H_d$ \cite{Weinberg-VIII}, where the waved fields
denote the basis of eigen states of mass matrix after taking into account for
the corrections to the symmetry of generation democracy.} due to a small
breaking of democratic symmetry in the Higgs sector, whereas the hierarchy is
also admissible as for the quarks and charged leptons. Therefore, two
lightest generations of Higgs doublets would split, so that the only
generation of Higgs particles would determine the physics at the electroweak
scale, while two other generations remain heavy with the Planckian and
sub-Plamckian masses. In this way, the evolution of bare masses to lower
scales for the heavy generations of Higgs doublets is frozen, at least, under
a sub-Planckian scale, whereat the heavy states have to decouple, hence,
their propagation can be neglected. Then, these heavy generations do not
acquire any non-zero vacuum expectations, i.e. they do not contribute to the
masses of gauge bosons in the SM. In contrast, the primary mass-term of the
lightest generation of Higgs particles will evolve up to the electroweak
scale, whereat the minimum of potential is shifted from instable zero field
to its vacuum expectation value of actual magnitude\footnote{The square of
mass at zero of Higgs field becomes negative.}.

It is interesting to note that we have to introduce another kind of symmetry
for the the color-triplet components of Higgs particles: the nondegenerate
matrix of generation masses of diagonal kind with the elements of the same
order of $M$, but without any mixing\footnote{A small mixing in the form of
perturbations with no breaking of nondegeneration is admissible, of course,
and it will remain the scheme with three superheavy generations to be valid,
again.}. This method lifts the problem of splitting between the triplet and
doublet components of Higgs field, because we can formulate quite the natural
mechanism for the splitting with the only scale of $M$. Indeed, for the
triplet higgsinos we get
\begin{equation}\label{mix-chiggs2}
    \mathscr{L}_\mathrm{\varkappa}=\lambda_\mathrm{\varkappa}\hskip-2.pt\cdot\hskip-2.pt
    (\varkappa^c_{1},\varkappa^c_{2},\varkappa^c_{3})\hskip-2.pt
    \left(
                  \begin{array}{ccc}
                    M & 0 & 0 \\
                    0 & M & 0 \\
                    0 & 0 & M \\
                  \end{array}
                \right)\hskip-4.pt
    \left(\hskip-2.pt
    \begin{array}{c}
      \varkappa_{1} \\[1mm]
      \varkappa_{2} \\[1mm]
      \varkappa_{3}
    \end{array} \hskip-2.pt
    \right),
\end{equation}
that corresponds to the analogous term for the scalar Higgs triplets.

Thus, the problem of scale hierarchy becomes closely related to the problem
of triplet-doublet splitting for the Higgs fields. Moreover, these problems
get the solution with the single dynamical scale $M$, and they are reduced to
the mechanism of small breaking of ``democratic'' symmetry of generations in
the Higgs sector, that is quite analogous to the problem on the mechanism of
breaking of the ``democratic'' symmetry of generations in the Yukawa sector
of theory.

In addition, we have to emphasize that two kinds of mass-matrices:
non-generate and ``democratic'' ones, suggest the explicit breaking of
SU(5)-symmetry in the Yukawa sector of theory. In this way, the scale of such
breaking is Planckian, i.e. it is higher that $\Lambda$.

The same idea could be useful for the decoupling of mirror matter: in the
mirror Higgs sector the mixing is determined by a nondegenerate diagonal
matrix with the sterile scale $M\sim m_\mathrm{Pl}$, hence, all of Higgs
particles and fermions with the Dirac mass in the mirror world are
supermassive. Consequently, the breaking of electroweak symmetry is decoupled
from the the mirror world with the scale of $M$. The masses of mirror
neutrinos are Planckian, too (see Appendix \ref{appendix}).

\subsection{The supersymmetry}
The unification of fermionic components for chiral superfields of matter and
conjugated components of mirror matter into the common multiplet with the
Majorana components of real gauge superfields points to the fact that the
massless particles with the spin of $1$, $\frac12$ and $0$  enter the same
superfield of $E_8$, that assumes two steps with increment of $\frac12$
within the representation of supermultiplet, i.e. the supersymmetry of $N=2$.
However, this extended supersymmetry is not important for our consideration
of fermionic multiplet, hence, it can be broken at a Planckian scale.

\section{Conclusion}

Thus, we have shown how one can logically arrive to the unique variant of
ultimate supersymmetric theory of grand unification with the group $E_6$ for
the fundamental $\boldsymbol{27}$-plet of single generation of fermionic
matter and higgsinos, as well as with the group $E_8$ for the fundamental
$\boldsymbol{248}$-plet of three generations of fermions and gauginos.

Of course, the consideration of those groups themselves is not new in GUT:
see original articles \cite{E6-2,King1,King2} and the comprehensive review on
$E_6$ in \cite{E6}, as well as the papers on $E_8$
\cite{Bars:1980mb,Braun:2005bw}. We especially note article \cite{Reut1},
where the supersymmetric $E_6$ GUT is built in 2 steps by unifying two
coupling constants at one scale $\Lambda$ and further reaching the final
unification of third and combined coupling constant at another scale $M\gg
\Lambda$, that is similar to our approach. This mechanism produces a new
insight on the doublet-triplet splitting problem \cite{Reut1,Reut2}. However,
our construction allows us not only uniquely to substantiate the set of
irreducible representation in explicit nickname-components, but to introduce
the natural mechanism for generating both the hierarchy of scales and the
decoupling of superheavy states. This mechanism is certainly differs from the
approaches on the market: the mechanism by Dimopoulos and Wilczek
\cite{DimWil}, the pseudosymmetry with the Nambu-Goldstone boson
\cite{Barbieri:1992yy,Randall:1995sh} etc. \cite{Giudice:1988yz,Kim:1994eu}.

Let us emphasize that the offered schemes for the mass-matrices are
realistic, but for their consistent substantiation certainly requires to
write down both a Yukawa sector of interactions and a spectrum of vacuum
expectations values for the scalar fields in the explicit form as well as to
specify the potentials of self-action for these field, that result in the
spontaneous breaking of symmetry. These items are predominant directions for
the further development of our model.

It is interesting to note that the algebras of isometries in the projective
spaces of real $\mathbb R$ and complex $\mathbb C$ numbers, as well as
quaternions $\mathbb H$ are isomorphic to the algebras of generators for the
infinite series of classical simple compact groups of special orthogonal,
unitary and symplectic transformations \cite{Baez}:
\begin{equation}\label{isom-class}
\begin{array}{l}
    \mathfrak{isom}(\mathbb{RP}^n)\cong\mathfrak{so}(n+1),\\[2mm]
    \mathfrak{isom}(\mathbb{CP}^n)\cong\mathfrak{su}(n+1),\\[2mm]
    \mathfrak{isom}(\mathbb{HP}^n)\cong\mathfrak{sp}(n+1),
\end{array}
\end{equation}
while the exceptional groups $E_k$ are related to octonions $\mathbb O$, so
that there is the isomorphism for the isometries of projective planes
\cite{Baez}:
\begin{equation}\label{isom}
\begin{array}{l}
    E_6\cong \mathrm{Isom}((\mathbb C\otimes \mathbb O)\mathbb P^2),\\[2mm]
    E_7\cong \mathrm{Isom}((\mathbb H\otimes \mathbb O)\mathbb P^2),\\[2mm]
    E_8\cong \mathrm{Isom}((\mathbb O\otimes \mathbb O)\mathbb P^2),
\end{array}
\end{equation}
and the number of such the groups are finite because of the non-associativity
of octonion algebra. Thus, the ultimate group of GUT is mathematically based
on the maximal generalization of numbers, i.e. octonions.

The octonions are the non-associative generalization of quaternions, which
are widely used in the representation of hermitian matrices of $2\times 2$
for the description of Lorentz symmetry of Minkowskian space-time. In
addition, there are the following isomorphisms for the algebras of special
linear transformations of two-component vectors:
\begin{equation}\label{isom-sl}
    \begin{array}{l}
       \mathfrak{sl}(2,\mathbb C)\cong \mathfrak{so}(3,1), \\[2mm]
       \mathfrak{sl}(2,\mathbb H)\cong \mathfrak{so}(5,1), \\[2mm]
       \mathfrak{sl}(2,\mathbb O)\cong \mathfrak{so}(9,1), %%\\[2mm]
     \end{array}
\end{equation}
that probably points to a relation of ultimate group of GUT to the
10-dimensional space-time, which naturally appears in the theory of
superstrings. In this respect, the relation of two ways for constructing the
GUT becomes more predictable: the first way is given by the phenomenology of
local quantum field theory as we have done in this paper, while the second
way originates from the F-theory in superstrings \cite{Heckman:2010bq}. These
ways lead to the common result, i.e. the group $E_8$. However, remember that
the multiplet constructed in this paper contains the fields with the masses
of Planckian scale, that essentially differs from the multiplets in the
models inspired by superstrings.

This work was partially supported by the grant of Russian Foundations for
Basic Research 10-02-00061, the grant of Special Federal Program ``Scientific
and academics personnel'' for the Scientific and Educational Center
2009-1.1-125-055-008.

\appendix
\section{The mixing of mirror neutrinos\label{appendix}}
In the case of mirror neutrinos: the doublet, singlet and sterile particles,
the mixing matrix takes the following form up to the overall Yukawa factor:
\begin{equation}\label{M-nu-mirror}
    \mathscr{M}_\mathrm{mir.}=\left(
                  \begin{array}{ccc}
                    0 & M^\prime & 0 \\
                    M^\prime & 0 & \Lambda \\
                    0 & \Lambda & M \\
                  \end{array}
                \right),
\end{equation}
where $M^\prime\sim M\sim m_\mathrm{Pl}$. Since $\det
\mathscr{M}_\mathrm{mir.}=-M{M^\prime}^2$, this matrix is not degenerate, and
we expect that all of massive states have got the same scale of mass of the
order of $M$. Indeed, let us start with the rotation of two states under the
matrix
$$
    U_{12}^\mathrm{mir.}=
    \left(
      \begin{array}{ccc}
        c_{12}^\mathrm{mir.} & s_{12}^\mathrm{mir.} & 0 \\
        -s_{12}^\mathrm{mir.} & c_{12}^\mathrm{mir.} & 0 \\
        0 & 0 & 1 \\
      \end{array}
    \right),
$$
where $c_{12}^\mathrm{mir.}=s_{12}^\mathrm{mir.}=1/\sqrt{2}$, so that
$$
    {U^\mathrm{mir.}_{12}}^\dagger\cdot\mathscr{M}_\mathrm{mir.}\cdot
    U_{12}^\mathrm{mir.}=
    \left(
      \begin{array}{ccc}
        -M' & 0 & -\frac{\Lambda}{\sqrt{2}} \\[2mm]
        0 & M' & \frac{\Lambda}{\sqrt{2}} \\[2mm]
        -\frac{\Lambda}{\sqrt{2}} & \frac{\Lambda}{\sqrt{2}} & M \\
      \end{array}
    \right),
$$
and further,  we make the rotation by the angle $\theta_{23}^\mathrm{mir.}$,
given by the condition
$$
    \tan 2\theta_{23}^\mathrm{mir.}=\frac{\Lambda\sqrt{2}}{M-M'},
$$
after which the mass matrix of mirror neutrinos
$\tilde{\mathscr{M}}_\mathrm{mir.}={U_{23}^\mathrm{mir.}}^\dagger
{U_{12}^\mathrm{mir.}}^\dagger\cdot\mathscr{M}\cdot U_{12}^\mathrm{mir.}
U_{23}^\mathrm{mir.}$ gets the form
\begin{equation}\label{AM}
    \tilde{\mathscr{M}}_\mathrm{mir.}=
    \left(
      \begin{array}{ccc}
        -M' & \Lambda\,\frac{s_{23}^\mathrm{mir.}}{\sqrt{2}} &
        -\Lambda\,\frac{c_{23}^\mathrm{mir.}}{\sqrt{2}} \\[2mm]
        \Lambda\,\frac{s_{23}^\mathrm{mir.}}{\sqrt{2}} & M_2-\tilde\Lambda & 0 \\[2mm]
        -\Lambda\,\frac{c_{23}^\mathrm{mir.}}{\sqrt{2}} & 0 & M_3+\tilde\Lambda \\
      \end{array}
    \right),
\end{equation}
where we denote
$$
\begin{array}{l}
\displaystyle
    M_2=M'(c_{23}^\mathrm{mir.})^2+M(s_{23}^\mathrm{mir.})^2,\\[1mm]
\displaystyle
    M_3=M'(s_{23}^\mathrm{mir.})^2+M(c_{23}^\mathrm{mir.})^2,\\[1mm]
\displaystyle
    \tilde\Lambda=\Lambda\,\frac{\sin2\theta_{23}^\mathrm{mir.}}{\sqrt{2}}.
\end{array}
$$
The perturbation theory is applicable to matrix (\ref{AM}), so that the
mirror neutrinos get the masses
\begin{equation}\label{Mirr}
\begin{array}{l}
\displaystyle
    m_1^\mathrm{mir.}\approx
    -M'-\frac{\Lambda^2(s_{23}^\mathrm{mir.})^2}{2(M'+M_2)}
    -\frac{\Lambda^2(c_{23}^\mathrm{mir.})^2}{2(M'+M_3)},\\[4mm]
\displaystyle
    m_2^\mathrm{mir.}\approx M_2-\tilde\Lambda
    +\frac{\Lambda^2(s_{23}^\mathrm{mir.})^2}{2(M'+M_2)},\\[4mm]
\displaystyle
    m_3^\mathrm{mir.}\approx M_2+\tilde\Lambda
    +\frac{\Lambda^2(c_{23}^\mathrm{mir.})^2}{2(M'+M_3)}.%%%\\[mm]
\end{array}
\end{equation}
As we have expected these values have got the same order of magnitude of
Planckian scale. At this scale the breaking of lepton number for the mirror
neutrinos becomes essential due to the mixing of initial states as given by
both the rotations described above and the corrections in the perturbation
theory. but we do not consider the mixing here.

\end{document}